# Exploration of Zeolites as High-Performance Electrode Protective Layers for Alkali-Metal Batteries

Lina Wang[1,2,3] and Guangfu Luo[4,5*]

[1]*Fujian Science & Technology Innovation Laboratory for Optoelectronic Information of China, Fuzhou, Fujian 350108, China*

[2]*Fujian Key Laboratory of Green Extraction and High-Value Utilization of New Energy Metals, Fuzhou University, Fuzhou, Fujian 350108, China*

[3]*State Key Laboratory of Structural Chemistry, and Fujian Provincial Key Laboratory of Materials and Techniques toward Hydrogen Energy, Fujian Institute of Research on the Structure of Matter, Chinese Academy of Sciences, Fuzhou, Fujian 350002, China*

[4]*Department of Materials Science and Engineering, Southern University of Science and Technology, Shenzhen 518055, China*

[5]*Guangdong Provincial Key Laboratory of Computational Science and Material Design, Southern University of Science and Technology, Shenzhen 518055, China*

**E-mail: luogf@sustech.edu.cn*

## Abstract

The electrode-electrolyte interfaces play pivotal roles in alkali-metal batteries, necessitating superior electrochemical stability, excellent electrical insulation, and high ionic conductivity. This study proposes using zeolites as interfacial protective layers owing to their inherently high stability with both alkali metals and high-voltage cathodes, as well as exceptionally wide bandgaps that minimize electron transport. To further pinpoint zeolites with rapid ionic diffusivity among their versatile structures, we devise a universal approach to explore diffusion dynamics in arbitrary structures. Through first-principles calculations, we identify the diffusion networks of $Li^+$, $Na^+$, and $K^+$ in twenty-two, seventeen, and four zeolites, respectively. Eventually, we predict five, seven, and three zeolites as suitable interfacial protective layer for lithium-, sodium-, and potassium-metal batteries, respectively, each characterized by a diffusion barrier below 0.3 eV. This research automates the exploration of diffusion dynamics in complex materials and underscores the significant potential of zeolites as interfacial protective layers in alkali-metal batteries.

## Introduction

Alkali-metal batteries, encompassing lithium, sodium, and potassium variants, hold immense potential to revolutionize energy storage due to the low redox potentials of alkali metal anodes, and their energy densities can be further enhanced when paired with high-voltage cathodes[1,2]. However, the highly reductive alkali metal anodes and oxidative cathodes pose significant challenges in maintaining stability with either liquid or solid electrolytes[3-7], resulting in the spontaneous formation of solid-electrolyte interphase (SEI) and cathode-electrolyte interphase (CEI). The continuous generation of these interphases induce ongoing electrolyte loss, diminished Coulombic efficiency, dendrite formation, and eventually battery failure[8-10]. Versatile strategies have been proposed to address these challenges, such as optimizing the electrolyte components[11,12], incorporating porous interphases on electrodes to facilitate uniform ion nucleation[13,14], and integrating alkali-plating hosts to alleviate the stress[15,16].

Current interfacial materials face significant challenges in simultaneously fulfilling the stringent requirements for electrochemical stability towards alkali metals and cathodes, rapid ion diffusivity, low electronic conductivity, and affordability[17-20]. The electrochemical stability is influenced by the relative positions of interfacial materials' band edges to the electrochemical potentials of electrodes[21-24] and subsequent chemical reactions[25]. Specifically, a stable material should exhibit a conduction band minimum (CBM) energy higher than the electrochemical potential of $M^+/M$ to avoid reduction by alkali metal M, and a valence band maximum (VBM) energy lower than the CBMs of cathodes to prevent oxidation. Figure 1a illustrates that only a few solids and electrolytes meet these stability criteria. Notably, AlN, $Al_2O_3$, and $SiO_2$ stand out with high CBMs of -1.3, -1.0, and -0.9 eV, and low VBMs of -7.4, -9.8, and -9.9 eV, respectively. Their exceptionally wide band gaps also ensure excellent electronic insulation. $SiO_2$, particularly when in the form of zeolites, has been reported recently with the exceptional stability, where a Faujasite (FAU)-type zeolite membrane was used as a solid electrolyte in rechargeable lithium-metal batteries[26]. The zeolite exhibits crystalline particle sizes of about 28 μm and demonstrates remarkable stability towards lithium-metal anodes. Moreover, zeolites demonstrate cost-effectiveness and versatile structures with uniform nanopores, which facilitate the migration of ions and molecules along their channels and cavities[27,28], making them promising candidates as high-performance interphase layers for alkali-metal batteries.

Previous efforts have been made in the prediction of ionic transport based on the bond valence method[29-32], effective medium theory[33-35], random resistance model[37], and subsequent machine learning techniques[36,37]. However, revealing the diffusion dynamics of alkali ions in zeolites based on first-principles calculations poses significant challenges. *Ab initio* molecular dynamics simulations are often limited by time scale constraints, which hinders their ability to predict large migration barriers. While first-principles transition state calculations can assess process with versatile migration energy barriers, they necessitate identifying stable adsorption sites and potential diffusion paths, which is particularly complex for intricate zeolite structures.

In this study, we advocate for the utilization of zeolites as promising interfacial protective layers for alkali-metal batteries due to their intrinsic electronic insulation and excellent chemical stability with alkali metal anodes and all high-voltage cathodes. To ascertain the diffusivities of $Li^+$, $Na^+$, and $K^+$ within these intricate nanoporous structures, we introduce a universal approach for automatically searching diffusion paths in arbitrary structures and eventually identity five favorable zeolites for Li-metal batteries, seven for Na-metal batteries, and three for K-metal batteries.

## Results and Discussion

### Charge Transfer Stability between Zeolites and Alkali Metal Anodes and Cathodes

To comprehensively evaluate the electrochemical stability of a coating material, both the interfacial charge transfer and thermodynamics of potential interfacial reactions are often assessed[22-25,38,39]. In this study, the interfacial charge transfer stability of zeolites is evaluated by calculating the band edges of several representatives, namely, AFN, EDI, BCT, and ERI, based on first-principles calculations (see details in Methods). As depicted in Figure 1a, AFN, EDI, BCT, and ERI zeolites exhibit wide band gaps of 7.46, 7.26, 7.08, and 7.44 eV, respectively. Their VBMs are over 2.19 eV lower than the CBMs of all typical cathodes, including the high-voltage cathode $LiNi_{0.5}Mn_{1.5}O_4$, indicating strong resistance to oxidation by cathodes. Furthermore, their CBMs are 1.30, 1.53, 1.53, and 1.64 eV below the vacuum level, respectively, which are higher or slightly lower than the electrochemical potentials of $Li^+/Li^0$, $Na^+/Na^0$, $K^+/K^0$, suggesting resistance to reduction by alkali metals. These zeolites exhibit consistency in band gaps and band-edge positions, attributed to their identical composition of $SiO_2$ and shared coordination environments of Si and O. Additional analyses of density of states (DOS) for ERI before and after contact with Li metal, along with charge density difference, indicate that the bulk region of ERI retains its insulator nature after contact with Li (Figure 1b, c). Based on these results, we conclude that zeolites demonstrate not only low electronic conductivity due to their wide band gaps but also robust charge transfer stability when interfacing with alkali metal anodes and high-voltage cathodes.

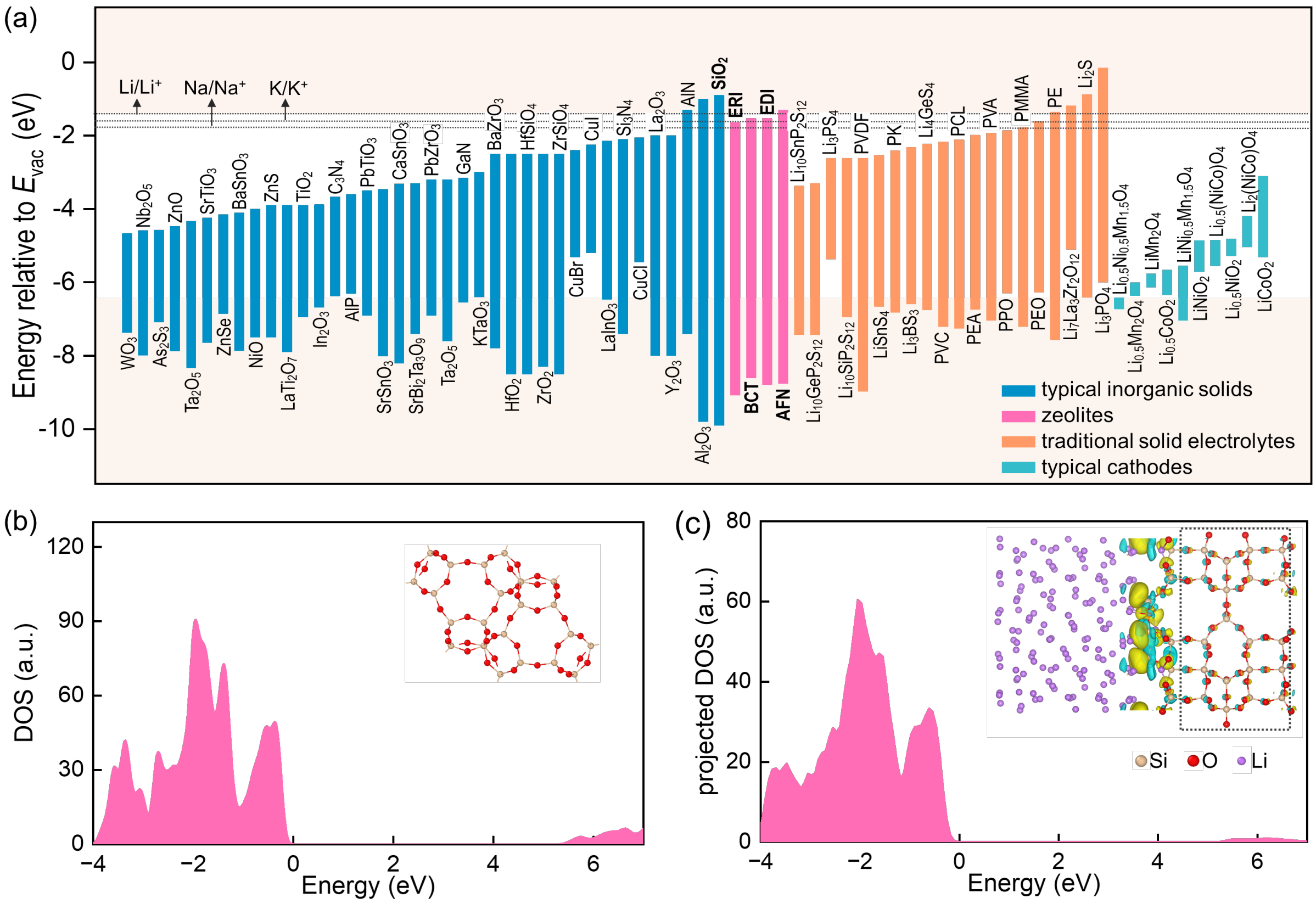

**Figure 1.** (a) Band edges of typical inorganic solids, cathodes, solid electrolytes from experiments[23,38,40-43], and those of several zeolites obtained by our first-principles calculations, alongside the electrochemical potentials of Li/Li$^+$, Na/Na$^+$, and K/K$^{+2}$. The vertical bars indicate the band gap regions, and the vacuum level is set to zero. The upper and lower shaded regions highlight the desirable ranges for the CBM and VBM of electrolytes that lead to potential inertness towards alkali-metal anodes and cathodes, respectively. (b) DOS and optimized structure of ERI bulk. (c) Projected DOS of ERI near the interface in the Li/ERI heterojunction (boxed region), as well as the charge density difference shown in the inset.

We also investigate the potential reactions between zeolites and electrodes to evaluate their stability under electrochemical conditions[25]. The formation enthalpy of Li-Si-O, Na-Si-O, and K-Si-O systems under electrochemical voltages are shown in Figure 2. The results indicate that zeolites, which are primarily composed of $SiO_2$, exhibit the highest stability against potential reaction products with anodes over the voltage range of 0–5 V versus $Li^+/Li^0$, $Na^+/Na^0$, or $K^+/K^0$. Since zeolites are fully oxidized, they are inherently resistant to further oxidation by cathodes. Notably, a recent experimental study demonstrated that a FAU-type zeolite membrane used as an electrolyte exhibited remarkable stability towards Li-metal anode[26]. Given the similar composition of zeolites, it is likely that other zeolites also exhibit high electrochemical stability.

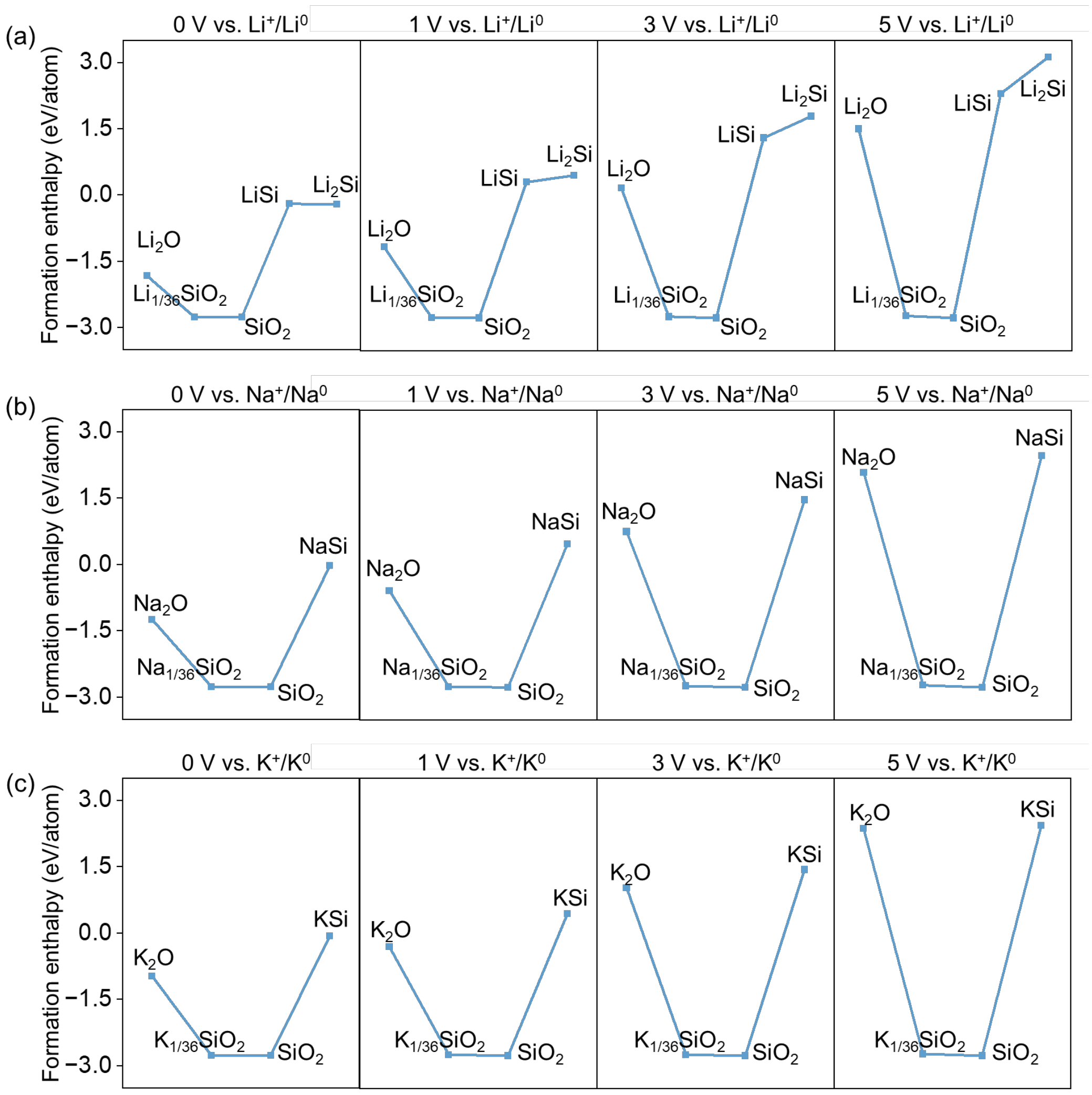

**Figure 2.** Energy convex hull of potential products formed between zeolites and the (a) Li, (b) Na, and (c) K metal anodes under electrochemical conditions. The reference states are Si bulk, $O_2$ molecule, and alkaline metals.

**Screening of Zeolites with Fast Alkali Metal Ion Transportation**

In order to identify zeolites capable of conducting alkali metal ions while impeding electrolyte permeation[13], we focuses on those with a maximum free-diffusion-sphere diameter in the range of [$r_{M+}$, $r_{electrolyte}$]. Here, $r_{M+}$ represents the diameter of alkali metal ion (1.80, 2.32, and 3.04 Å for $Li^+$, $Na^+$, and $K^+$, respectively) and $r_{electrolyte}$ is the diameter of the liquid electrolyte (~3.45 Å for EC). We further exclude systems with a total number of atoms exceeding 192 and a lattice constant over 35 Å, due to computational challenges. Eventually, from a pool of 251 candidates in the zeolite database[44], we identify twenty-two candidates for $Li^+$ diffusion, seventeen for $Na^+$ diffusion, and four for $K^+$ diffusion (see details of the selected zeolites in Section I of Supporting Information).

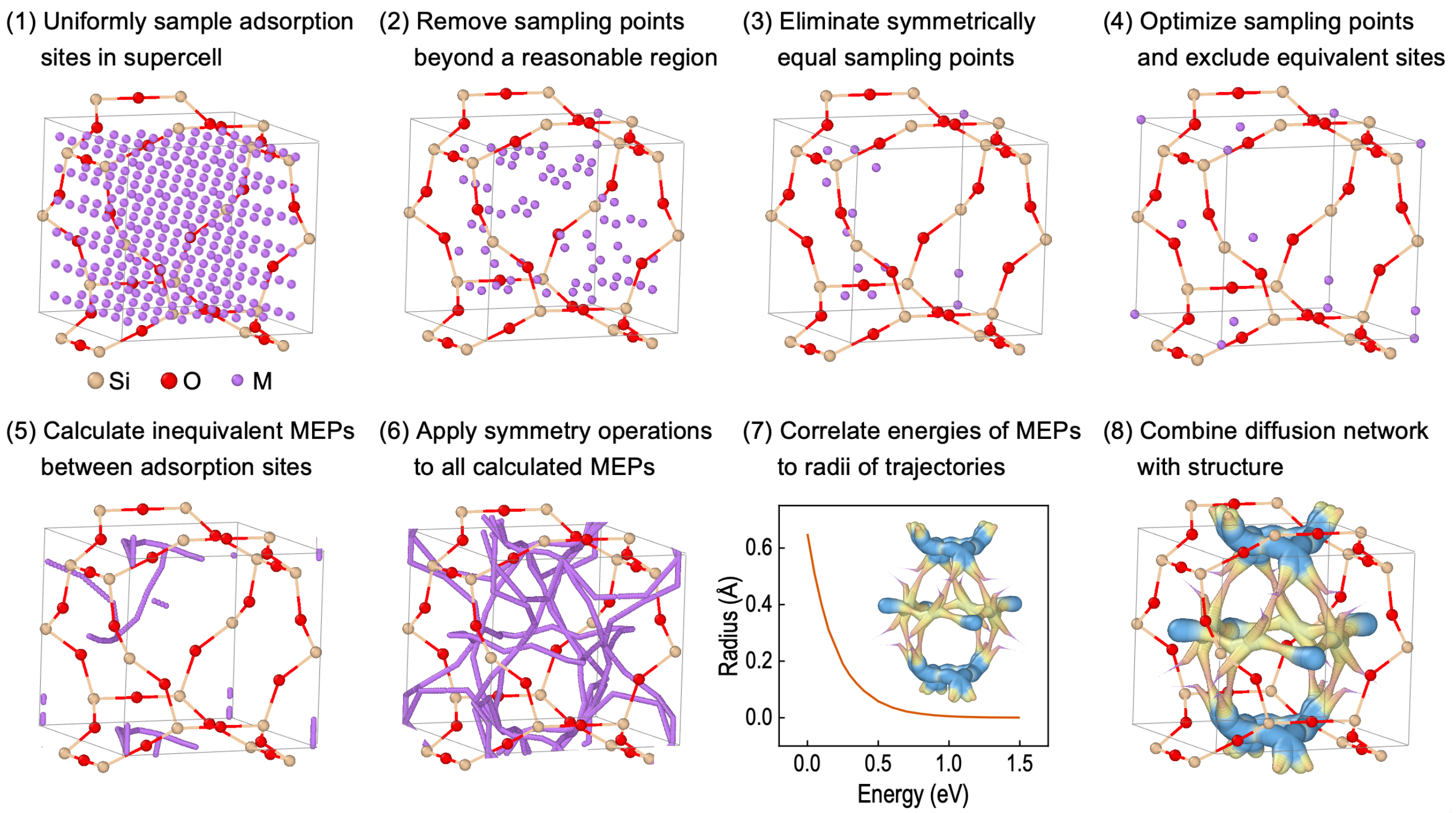

**Figure 3.** Flowchart of automatic search of 3D diffusion network for species M in a complex structure.

Given the complex structure of zeolites, we develop an automatic search approach to identify the 3D diffusion network of alkali metal ions within zeolites. As schematically illustrated in Figure 3, this approach consists of the following eight steps, which enable a comprehensive acquisition of the diffusion network in arbitrary structures.

Step 1: Uniformly sample the adsorption sites in a supercell, such as zeolite EDI in Figure 3. The sampling spacing can be adjusted according to specific cases and a value of 1.0 Å is used in this study.

Step 2: Remove all sampling points outside a reasonable distance range relative to the nearest atom, i.e., 1.8–2.2 Å for $Li^+$, 2.2–2.6 Å for $Na^+$, and 2.5–2.9 Å for $K^+$, given the typical bond lengths of 2.0, 2.4, and 2.7 Å for Li-O, Na-O, and K-O, respectively.

Step 3: Eliminate equivalent sampling points according to crystalline symmetry. A small symmetry tolerance of 0.2 Å is used here to account for the slight symmetry broken of sampling points.

Step 4: Perform first-principles calculations based on the density functional theory (DFT) to optimize the remaining inequivalent sampling points. Subsequently, exclude the equivalent optimized adsorption sites with a symmetry tolerance of 0.5 Å.

Step 5: Utilize crystalline symmetry to generate all the stable adsorption sites within the supercell based on the final adsorption sites obtained in step 4. Identify potential hopping events between all combinations of two adsorption sites, exclude the equivalent hoppings according to crystalline symmetry, discard those with hopping distances beyond a reasonable range (0.5–5 Å here) due to potentially too small or too large activation barriers, and eliminate hoppings that can be substitutable by shorter ones. Perform transition state calculations for all remaining hoppings to obtain the minimum energy paths (MEPs).

Step 6: Apply crystalline symmetry to the calculated MEPs to generate a 3D diffusion network.

Step 7: Generate an intuitive 3D diffusion network by correlating the energies along the MEPs with the radii of diffusion trajectories. In this study, the formula $r = 0.65\text{Å}\ e^{\frac{-E}{0.2\,\text{eV}}}$ is employed, where $r$ and $E$ are trajectory radius and energy, respectively. The diffusion network can also be colored to enhance the visualization.

Step 8: Integrate the 3D diffusion network in step 7 into the geometrical structure.

A detailed explanation of the chosen parameters is provided in Section II of the Supporting Information.

Using the aforementioned approach, we successfully unveil the diffusion networks of $Li^+$, $Na^+$, and $K^+$ within the selected zeolites (Section III of Supporting Information). Figure 4a illustrates the diffusion energy barriers of $Li^+$ across twenty-two zeolites, where the energy barriers range from 0.18 to 1.21 eV. Notably, twelve zeolites exhibit energy barriers less than 0.65 eV, a value akin to those observed in typical solid electrolytes[45]. The diffusion networks in Figure 4b and Figure S5 demonstrate that $Li^+$ can readily diffuse in three directions in most of the twelve zeolites, except for the ERI zeolite, where the $Li^+$ diffusion is predominantly one-dimensional. The top five ones with the lowest diffusion energy barriers are the BCT, LTJ, APC, NSI, and EDI zeolites, with corresponding barriers of 0.18, 0.27, 0.27, 0.29, and 0.29 eV, respectively.

In the case of $Na^+$ diffusion (Figure 5a and Figure S6), the diffusion energy barriers span from 0.14 to 0.94 eV across the seventeen zeolites, with thirteen exhibiting barriers below 0.65 eV. Their diffusion networks depicted in Figure 5b reveal that $Na^+$ diffusion is three-dimensional in most zeolites, but largely two-dimensional in the MAR and AFG zeolites. The top seven ones with the lowest diffusion energy barriers are the BCT, APC, EDI, AHT, NSI, GOO, and AFN zeolites, with respective barriers of 0.14, 0.19, 0.19, 0.23, 0.23, 0.27 eV and 0.29 eV.

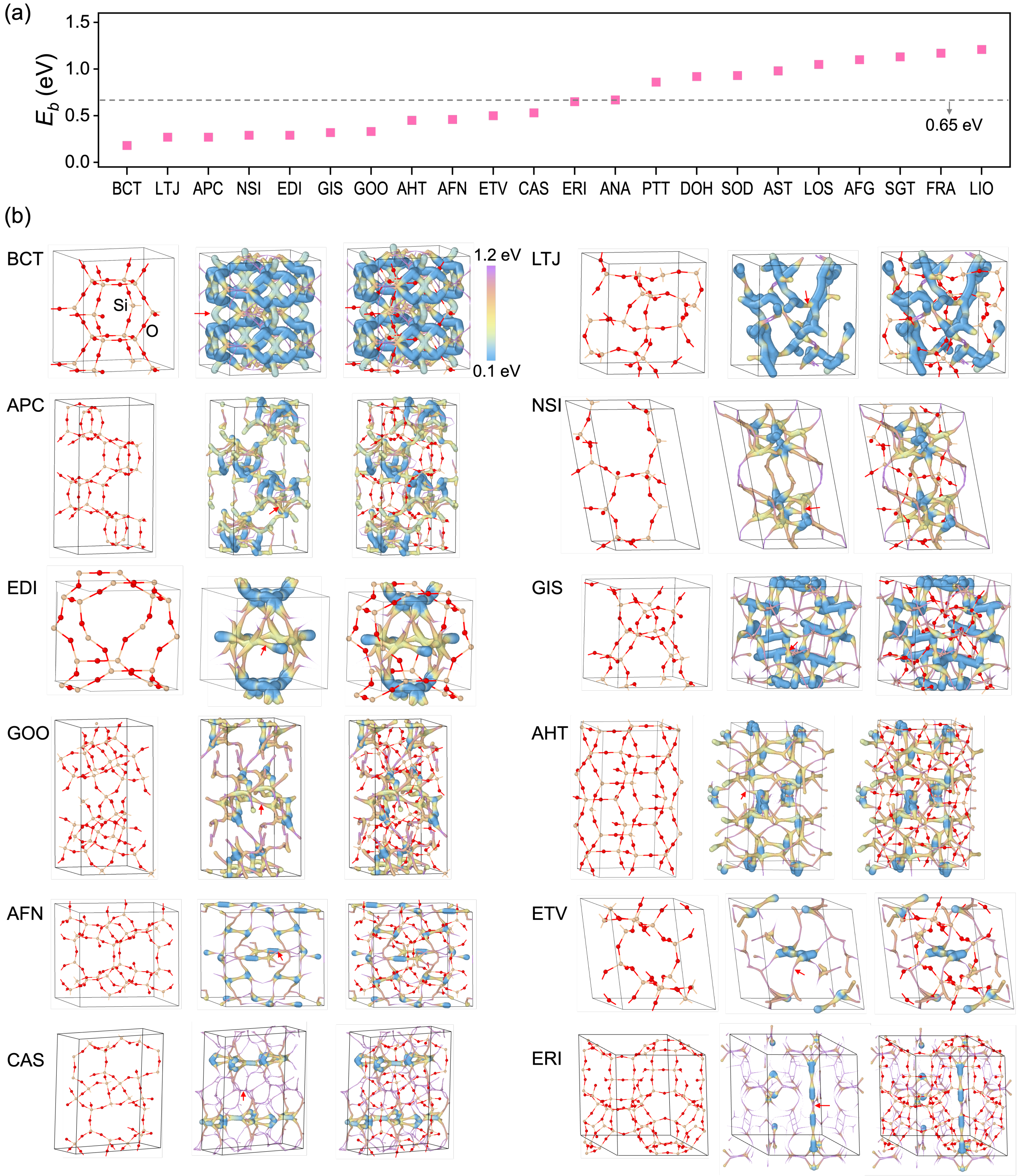

**Figure 4.** (a) Diffusion energy barrier $E_b$ of $Li^+$ in twenty-two zeolites, and (b) diffusion networks of $Li^+$ in zeolites with an energy barrier lower than 0.65 eV. The first to third columns depicts the zeolite structure, diffusion network, and their combination, respectively. The diffusion network is also colored to enhance the visualization. The diffusion-limiting sites are indicated by red arrows in the diffusion networks.

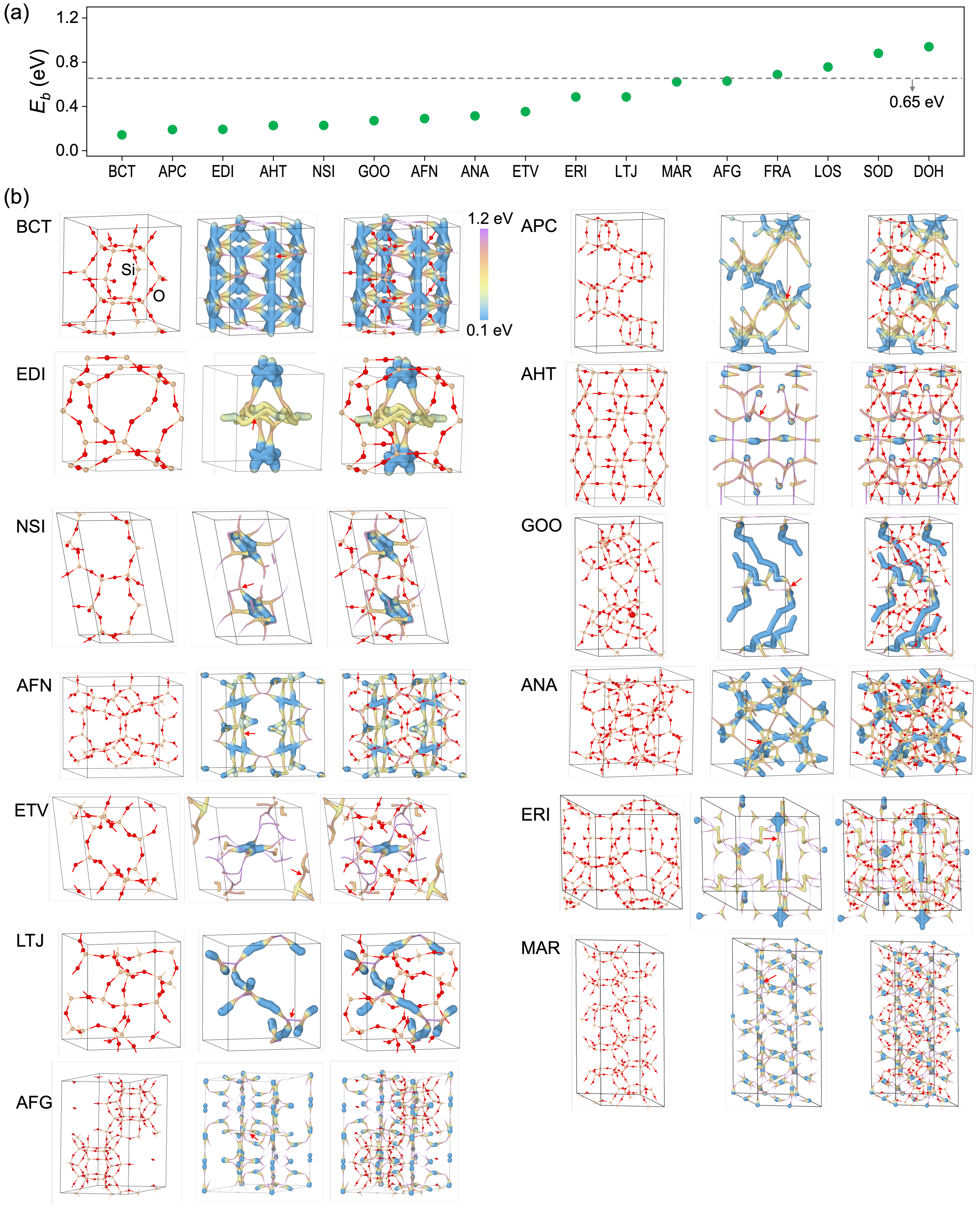

**Figure 5.** (a) Diffusion energy barrier $E_b$ of $Na^+$ in seventeen zeolites, and (b) diffusion networks of $Na^+$ in zeolites with an energy barrier lower than 0.65 eV. The first to third columns depicts the zeolite structure, diffusion network, and their combination, respectively. The diffusion-limiting sites are indicated by red arrows in the diffusion networks.

For $K^+$ diffusion, all the four examined zeolites exhibit diffusion energy barrier less than 0.65 eV (Figure 6a and Figure S7). Specifically, the LTJ zeolite possesses the lowest energy barrier of 0.11 eV, followed by 0.13 eV in GIS, 0.21 eV in EDI, and 0.61 eV in ERI. Compared to $Li^+$ and $Na^+$ diffusion in the same zeolites, the energy barriers for $K^+$ are noticeably lower, and the number of diffusion paths of $K^+$ is fewer (Figure 6). For instance, the diffusion energy barriers in the EDI zeolite are 0.29, 0.19, and 0.17 eV for $Li^+$, $Na^+$, and $K^+$, respectively. Further analyses indicate that the reduced number of diffusion paths is caused by the increasing cation radius from $Li^+$ to $Na^+$ and $K^+$, as leads to fewer stable adsorption sites. Moreover, larger cation ions involve a greater number of neighboring atoms along the diffusion path, consequently lowering the diffusion barrier. Taking the diffusion-limited step in the EDI zeolite as an example, the initial and transition states for $K^+$ diffusion feature four and two oxygen neighbors, respectively, whereas the corresponding structures for $Na^+$ and $Li^+$ possess only one neighboring oxygen atom (Figure S8 in Section IV of Supporting Information). Therefore, the distinct radii of $Li^+$, $Na^+$, and $K^+$ result in distinct adsorptions sites and diffusion paths within the nanopores of zeolites.

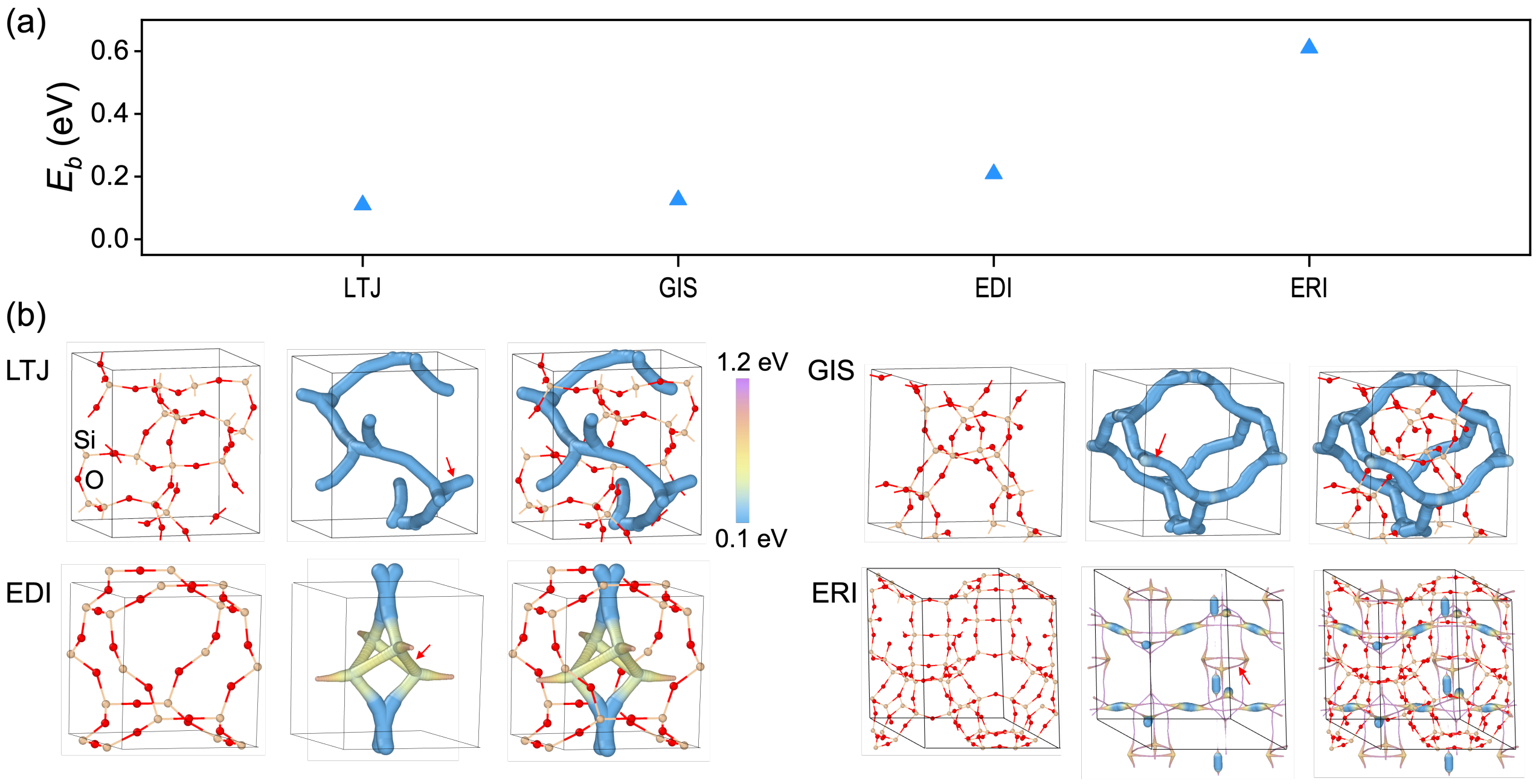

**Figure 6.** (a) Diffusion energy barrier $E_b$ of $K^+$ in four zeolites, and (b) diffusion networks in zeolites with an energy barrier lower than 0.65 eV. The first to third columns depicts the zeolite structure, diffusion network, and their combination, respectively. The diffusion-limiting sites are indicated by red arrows in the diffusion networks.

**Relationship between Diffusion Energy Barrier and Zeolite Geometry**

To elucidate the relationship between the diffusion energy barrier and zeolite gemometry, we examine two directly relevant states: transition state in the diffusion-limited step and the lowest adsorption site in the entire structure. As depicted in Figures S9-S10 (Section V of Supporting Information), the adsorption energy of both the lowest adsorption site and the transition state, $E_a^{min}$ and $E_a^{TS}$, follows a decending order for $Li^+$, $Na^+$, and $K^+$. To establish a relationship between the these adsorption energies and local structural features, such as the distances of cation-oxygen and cation-silicon, we employ an equation comprising a Coulombic interaction and a 12-10 Lennard-Jones formula for van der Waals interaction, as detailed in Section VI of Supporting Information. The fitting results in Figure 7a are generally satisfactory. Figure 7b and Figure S11 reveal that the cation-oxygen energy generally increases with distance, while the cation-silicon energy decreases with distance. In orther words, the cation-oxygen is attractive but the cation-silicon interaction is repulsive, aligning with the positive charge states of cations and silicon, and the nagative charge of oxygen.

Using $Li^+$ diffusion in the GOO and SOD zeolites as examples, which exhibit dramatically different diffusion barriers of 0.33 and 0.93 eV, respectively, we delve into the local structures of the diffusion-limited step. As shown in Figure 7c, the diffusion-limited step in the GOO zeolite occurs between two adjacent eight-membered rings with one sharing Si-O-Si-O-Si edge. In the transition state, $Li^+$ forms bonds with four neighboring oxygen atoms above the shared edge. Conversely, in the SOD zeolite (Figure 7d), the diffusion-limited step occurs between two adjacent six-membered rings with one sharing Si-O-Si edge. Here, in the transition state, $Li^+$ forms only one bond with a neighboring oxygen atom above the shared edge, suggesting a higher energy state than in the GOO zeolite. These structural features clealrly indicate that the the adsorption energy and ion transport are largely determined by the local interactions between alkali ion and neighboring oxygen atoms in zeolites.

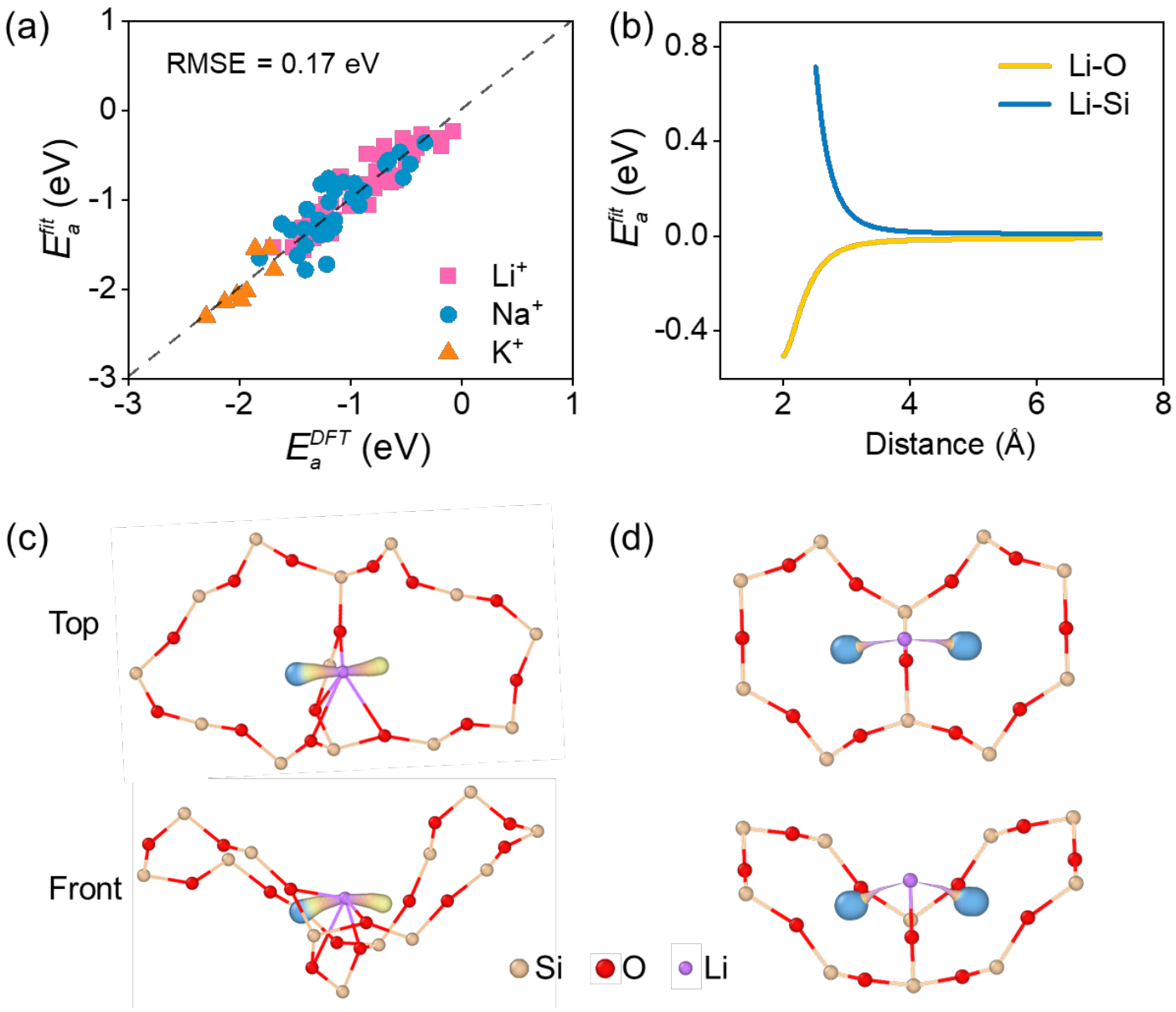

**Figure 7.** (a) Comparison of adsorption energy $E_a$ of all stable adsorption sites and transition state between fitting and our DFT results for $Li^+$, $Na^+$, and $K^+$ in all examined zeolites. (b) Fitting function of adsorption

energy with respect to Li-O and Li-Si distances in all examined zeolites. Local structures of the transition state in the diffusion-limited step for $Li^+$ diffusion in the (c) GOO and (d) SOD zeolites, together with the colored hopping paths.

Beyond the high stability and low diffusion energy barrier, the protective layer of electrodes should ideally exhibit other properties, such as low interfacial resistance, good crystallinity, and high mechanical strength[46,47]. While traditional seed growth methods[26] or hydrothermal reaction techniques[48,49] have been employed to grow zeolite membranes, alternative techniques, such as in situ growth on lithium metal or sputtering followed by post-annealing, may offer advantages by reducing interface impedance, improving adhesion[26], and producing highly uniform thin films[50,51]. Significant efforts have also been made to obtain large-particle-size zeolites with good crystallinity, such as the successful synthesis of ~28 μm FAU zeolite[26] and ~23 μm SSZ-13 zeolite[52]. Regarding the mechanical properties, the Young's moduli of zeolites range from 11.5 to 114.5 GPa[53,54], which are significantly higher than those of bulk alkali metals (~4.9 GPa for Li[55,56], ~2.3 GPa for Na[57,58], and ~1.3 GPa for K[59]). According to the Monroe-Newman theory, an interfacial layer with Young's modulus greater than that of the alkali metal can help inhibit dendrite growth by reducing the roughness of the alkali metal surfaces[56,59,60].

## Conclusion

In summary, we computationally investigate zeolites with high alkali-ion diffusivities among these inherently stable and electronically insulating materials, aiming to leverage them as electrode protective layer in alkali-metal batteries. We develop a general approach for automatically exploring diffusion networks in arbitary structures, leading to the indentification of twelve, thirteen, and four zeolites with diffusion barriers less than 0.65 eV for $Li^+$, $Na^+$, and $K^+$, respectively. Notably, the foremost zeolites dislaying diffusion barriers below 0.3 eV include BCT, LTJ, APC, NSI, and EDI for $Li^+$; BCT, APC, EDI, AHT, NSI, GOO, and AFN for $Na^+$; and LTJ, GIS, and EDI for $K^+$. Our findings suggest that the relative sizes of alkali metal ions and the pore size of zeolites significantly impact the adsorption sites and diffusion barriers. A closer proximity of alkali metal ions to oxygen atoms correlates with lower diffusion barriers. Given their robust electronic insulation and exceptional stability, zeolites with fast ionic diffusivities are favorable for use as protective layers in rechargable alkali-metal batteries.

## Methods

**Computational Details.** The first-principles calculations in this work are conducted based on the density functional theory, utilizing the Vienna *Ab-initio* Simulation Package[61]. The generalized gradient approximation in the Perdew-Burke-Ernzerhof form[62] is selected as the exchange-correlation functional for most calculations, except for the band edges, which are calculated using the hybrid exchange-correlation functional, HSE06[63], for more accurate description. A plane-wave energy cutoff of 521 eV is chosen, along with the following projector augmented wave pseudopotentials[64]: Si_GW ($3s^2 3p^2$) for silicon, O_GW ($2s^2 2p^4$) for oxygen, Li_GW ($2s^1$) for lithium, Na_PBE ($3s^1$) for sodium, and K_pv ($3p^6 4s^1$) for potassium. Monkhorst-Pack meshes with a *k*-point spacing of ~2π/30 $Å^{-1}$ are employed for the sampling of Brillouin zones. The convergence tolerance for energy and force in geometry optimization are $10^{-5}$ eV and 0.01 eV $Å^{-1}$, respectively. An adaptive semi-rigid body approximation (ASRBA) method[65] is used to construct the initial guess for transition state calculations, which are realized using the climibing nudged elastic band method[66]. The program for automatic search approach to identify the 3D diffusion network in this work is writtern using the Mathematica software[67]. The zeolite structures studied are sourced from the zeolite database[44], with the lattice constants and other relevant information provided in Table S1. The diffusion energy barrier ($E_b$) for the diffusion-limited step of alkali ion in zeolite is defined as Equation 1. The adsorption energy of transition state and lowest adsorption site, $E_a^{TS}$ and $E_a^{min}$, are defined as Equation 2 and Equation 3, respectively. The chemical potential of alkali metal ions is defined as Equation 4:

$$E_b = E_a^{TS} - E_a^{min} \tag{1}$$

$$E_a^{min} = E_{zeolite+M}^{min} + E_{FNV}^{min} - \mu_M - E_{zeolite} + q(E_{zeolite}^{VBM} + E_F) \tag{2}$$

$$E_a^{TS} = E_{zeolite+M}^{TS} + E_{FNV}^{TS} - \mu_M - E_{zeolite} + q(E_{zeolite}^{VBM} + E_F) \tag{3}$$

$$\mu_M = E_M \tag{4}$$

Here, $E_{zeolite+M}^{min}$ and $E_{zeolite+M}^{TS}$ are the total energies of zeolite adsorbed with alkali ion $M$ ($M$ = $Li^+$, $Na^+$, and $K^+$) at the lowest adsorption site and the transition state in the diffusion-limited step, respectively. Meanwhile, $E_{FNV}^{min}$ and $E_{FNV}^{TS}$ are corrections[68] for charged states in periodic systems for the lowest adsorption site and the transition state, respectively. $E_{zeolite}$ is the total energy of zeolite, $q$ is the net charge of alkali metal ion, namely +1; $E_{zeolite}^{VBM}$ and $E_F$ are the valence band maximum (VBM) and the Fermi level of zeolite; $E_M$ is the total energy per atom in bulk alkali metal.

**Fitting between Adsorption Energies and Local Structural Features.** We employ Equation 5 to describe the relationship between adsorption energy $E_a$ (Figures S9 and S10) and distances of M-O and M-Si, denoted as $r_{M\text{-O}}$ and $r_{M\text{-Si}}$, with consideration of the Coulombic interaction and a 12-10 Lennard-Jones expression for van der Waals force. The coefficients are determined using the least squares method, with the respective values in Table 1. The root mean square fitting errors are 0.149, 0.216, and 0.147 eV for $Li^+$, $Na^+$, and $K^+$ diffusion, respectively. The fitting curves of $Li^+$, $Na^+$, and $K^+$ are plotted in Figure S11.

$$E_a = \sum_{r_{\text{M-O}}} \left( \frac{a_1}{r_{\text{M-O}}} + \frac{a_2}{r_{\text{M-O}}^{12}} - \frac{a_3}{r_{\text{M-O}}^{10}} \right) + \sum_{r_{\text{M-Si}}} \left( \frac{b_1}{r_{\text{M-Si}}} + \frac{b_2}{r_{\text{M-Si}}^{12}} - \frac{b_3}{r_{\text{M-Si}}^{10}} \right) \tag{5}$$

Table 1. Fitting parameters in Equation 5 for $Li^+$, $Na^+$, and $K^+$.

| | $a_1$ | $a_2$ | $a_3$ | $b_1$ | $b_2$ | $b_3$ |
|---|---|---|---|---|---|---|
| $Li^+$ | $-5.88 \times 10^{-2}$ | $9.42 \times 10^{3}$ | $2.84 \times 10^{3}$ | $6.14 \times 10^{-2}$ | $2.84 \times 10^{4}$ | $-2.31 \times 10^{3}$ |
| $Na^+$ | $3.10 \times 10^{-2}$ | $6.12 \times 10^{4}$ | $1.34 \times 10^{4}$ | $1.01 \times 10^{-1}$ | $6.51 \times 10^{5}$ | $4.49 \times 10^{4}$ |
| $K^+$ | -3.22 | $5.48 \times 10^{5}$ | $6.15 \times 10^{4}$ | 6.52 | $5.97 \times 10^{6}$ | $5.48 \times 10^{5}$ |

**Supporting Information**

The Supporting Information is available free of charge on the ACS Publications website; Zeolites selected for investigation of alkali-metal ions diffusion; Parameters used in automatic search of 3D diffusion network; Diffusion energy barriers and networks of $Li^+$, $Na^+$, and $K^+$ in examined Zeolites; Comparison of adsorption sites and diffusion pathways of $Li^+$, $Na^+$, and $K^+$ in Zeolite EDI; Adsorption energy of the lowest adsorption site and that of the transition state in diffusion-limited step; Fitting between adsorption energies and local structural features (PDF)

Mathematica_codes.zip (ZIP)

3D-diffusion-networks.zip (ZIP)

**Acknowledgements**

This work was financially supported by the National Foundation of Natural Science, China (No. 52273226), Guangdong Provincial Key Laboratory of Computational Science and Material Design (Grant No. 2019B030301001), the Shenzhen Science and Technology Innovation Commission (No. JCYJ20200109141412308), the Postdoctoral Fellowship Program of CPSF, China (GZB20230758), China Postdoctoral Science Foundation (2024M753237), and Fujian Key Laboratory of Green Extraction and High-value Utilization of New Energy Metals (No. 2023-KFKT-2). The calculations were carried out on the Taiyi cluster supported by the Center for Computational Science and Engineering of Southern University of Science and Technology and also on the Major Science and Technology Infrastructure Project of Material Genome Big-science Facilities Platform supported by Municipal Development and Reform Commission of Shenzhen.

**Supporting Information**

Exploration of Zeolites as High-Performance Electrode Protective Layers for Alkali-Metal Batteries

Lina Wang[1,2,3] and Guangfu Luo[4,5*]

[1]*Fujian Science & Technology Innovation Laboratory for Optoelectronic Information of China, Fuzhou, Fujian 350108, China*

[2]*Fujian Key Laboratory of Green Extraction and High-Value Utilization of New Energy Metals, Fuzhou University, Fuzhou, Fujian 350108, China*

[3]*State Key Laboratory of Structural Chemistry, and Fujian Provincial Key Laboratory of Materials and Techniques toward Hydrogen Energy, Fujian Institute of Research on the Structure of Matter, Chinese Academy of Sciences, Fuzhou, Fujian 350002, China*

[4]*Department of Materials Science and Engineering, Southern University of Science and Technology, Shenzhen 518055, China*

[5]*Guangdong Provincial Key Laboratory of Computational Science and Material Design, Southern University of Science and Technology, Shenzhen 518055, China*

**E-mail: luogf@sustech.edu.cn*

## I. Zeolites Selected for Investigation of Alkali-Metal Ions Diffusion

**Table S1.** Zeolites with a maximum free-diffusion-sphere diameter in the range of [1.80 Å, 3.45 Å], [2.32 Å, 3.45 Å], and [3.04 Å, 3.45 Å], which are suitable for the diffusion of $Li^+$, $Na^+$, and $K^+$, respectively. Systems with atoms more than 192 and a lattice constant over 35 Å are excluded due to computational challenges. Initial structures are obtained from the database[1] approved by the Structure Commission of the International Zeolite Association.

| Short name | Full name | Chemical formula | Lattice constant (Å) | Maximum diameter of free sphere (Å) | | | Diffusion Ions | | |
|---|---|---|---|---|---|---|---|---|---|
| | | | | *a* | *b* | *c* | $Li^+$ | $Na^+$ | $K^+$ |
| AFG | Afghanite | $Si_{48}O_{96}$ | 12.5, 12.5, 20.8 | 2.55 | 2.55 | 2.54 | √ | √ | |
| AFN | AlPO-14 | $Si_{32}O_{64}$ | 14.0, 13.5, 10.2 | 2.85 | 2.53 | 2.85 | √ | √ | |
| AHT | AlPO-H2 | $Si_{24}O_{48}$ | 15.8, 9.2, 8.6 | 2.40 | 2.40 | 2.75 | √ | √ | |
| ANA | Analcime | $Si_{48}O_{96}$ | 13.6, 13.6, 13.6 | 2.43 | 2.43 | 2.43 | √ | √ | |
| ANO | AlPO-91 | $Si_{96}O_{192}$ | 12.9, 12.9, 40.4 | 3.26 | 3.26 | 1.84 | | | |
| APC | AlPO-C | $Si_{32}O_{64}$ | 9.0, 19.4, 10.4 | 3.16 | 2.57 | 2.6 | √ | √ | |
| AST | AlPO-16 | $Si_{40}O_{80}$ | 13.6, 13.6, 13.6 | 1.91 | 1.91 | 1.91 | √ | | |
| AVE | AlPO-78 | $Si_{48}O_{96}$ | 12.9, 12.9, 60.0 | 3.26 | 3.26 | 1.88 | | | |
| BCT | Mg-BCTT | $Si_{16}O_{32}$ | 9.0, 9.0, 5.3 | 2.55 | 2.55 | 2.91 | √ | √ | |
| CAS | Cesium Aluminosilicate | $Si_{24}O_{48}$ | 5.3, 14.1, 17.2 | 2.97 | 2.21 | 2.45 | √ | | |
| DOH | Dodecasil 1H | $Si_{34}O_{68}$ | 14.2, 14.2, 11.5 | 2.55 | 2.55 | 2.62 | √ | √ | |
| EDI | Edingtonite | $Si_{5}O_{10}$ | 6.9, 6.9, 6.4 | 3.20 | 3.20 | 3.44 | √ | √ | √ |
| EEI | ERS-18 | $Si_{200}O_{400}$ | 13.9, 35.7, 22.5 | 3.11 | 2.15 | 2.23 | | | |
| ERI | Erionite | $Si_{36}O_{72}$ | 13.1, 13.1, 15.2 | 3.42 | 3.42 | 3.42 | √ | √ | √ |
| ETV | EMM-37 | $Si_{14}O_{28}$ | 8.8, 9.6, 10.3 | 2.61 | 3.2 | 2.91 | √ | √ | |
| FAR | Farneseite | $Si_{84}O_{168}$ | 12.6, 12.6, 35.7 | 2.47 | 2.47 | 2.26 | | | |
| FRA | Franzinite | $Si_{60}O_{120}$ | 12.7, 12.7, 25.3 | 2.70 | 2.70 | 2.66 | √ | √ | |
| GIS | Gismondine | $Si_{16}O_{32}$ | 9.8, 9.8, 10.2 | 3.32 | 3.32 | 3.32 | √ | | √ |
| GIU | Giuseppettite | $Si_{96}O_{192}$ | 12.6, 12.6, 41.0 | 2.59 | 2.59 | 2.32 | | | |
| GOO | Goosecreekite | $Si_{32}O_{64}$ | 8.7, 11.0, 17.5 | 3.22 | 3.22 | 2.84 | √ | √ | |
| LIO | Liottite | $Si_{36}O_{72}$ | 12.3, 12.3, 15.6 | 2.37 | 2.37 | 2.29 | √ | | |
| LOS | Losod | $Si_{24}O_{48}$ | 12.6, 12.6, 10.3 | 2.56 | 2.56 | 2.56 | √ | √ | |
| LTJ | Linde Type J | $Si_{16}O_{32}$ | 9.3, 9.3, 10.1 | 3.12 | 3.12 | 3.12 | √ | √ | √ |
| LTN | Linde Type N | $Si_{768}O_{1536}$ | 35.6, 35.6, 35.6 | 2.08 | 2.08 | 2.08 | | | |
| MAR | Marinellite | $Si_{72}O_{144}$ | 12.4, 12.4, 30.7 | 2.43 | 2.43 | 2.43 | | √ | |
| MSO | MCM-61 | $Si_{90}O_{180}$ | 17.2, 17.2, 19.8 | 2.09 | 2.09 | 2.09 | | | |
| MTN | ZSM-39 | $Si1_{36}O_{272}$ | 19.9, 19.9, 19.9 | 2.61 | 2.61 | 2.61 | | | |
| NON | Nonasil | $Si_{88}O_{176}$ | 22.9, 15.7, 13.9 | 2.27 | 2.30 | 2.30 | | | |
| NSI | Nu-6(2) | $Si_{12}O_{24}$ | 14.1, 5.3, 8.9 | 2.60 | 3.30 | 2.45 | √ | √ | |
| PTT | PST-33 | $Si_{24}O_{48}$ | 12.9, 12.9, 10.0 | 3.02 | 3.02 | 1.90 | √ | | |
| SAT | STA-2 | $Si_{72}O_{144}$ | 12.9, 12.9, 30.6 | 3.25 | 3.25 | 3.25 | | | |
| SGT | Sigma-2 | $Si_{64}O_{128}$ | 10.3, 10.3, 34.4 | 2.11 | 2.11 | 2.11 | √ | | |
| SOD | Sodalite | $Si_{12}O_{24}$ | 9.0, 9.0, 9.0 | 2.53 | 2.53 | 2.53 | √ | √ | |
| SWY | STA-20 | $Si_{72}O_{144}$ | 13.1, 13.1, 30.3 | 3.43 | 3.43 | 3.43 | | | |
| TOL | Tounkite-like mineral | $Si_{72}O_{144}$ | 12.3, 12.3, 30.9 | 2.36 | 2.36 | 2.06 | | | |

## II. Parameters Used in Automatic Search of 3D Diffusion Network

### (1) Sampling spacing of adsorption sites in Step 1

To establish an appropriate sampling spacing for the adsorption sites, we compare the use of two values, 1.0 Å and 0.5 Å, in the BCT zeolite. As illustrated in Fig. S1, their 3D diffusion networks are largely unchanged and the overall diffusion energy barrier differs by only 0.04 eV. Therefore, we select a sampling spacing of 1.0 Å for all systems examined.

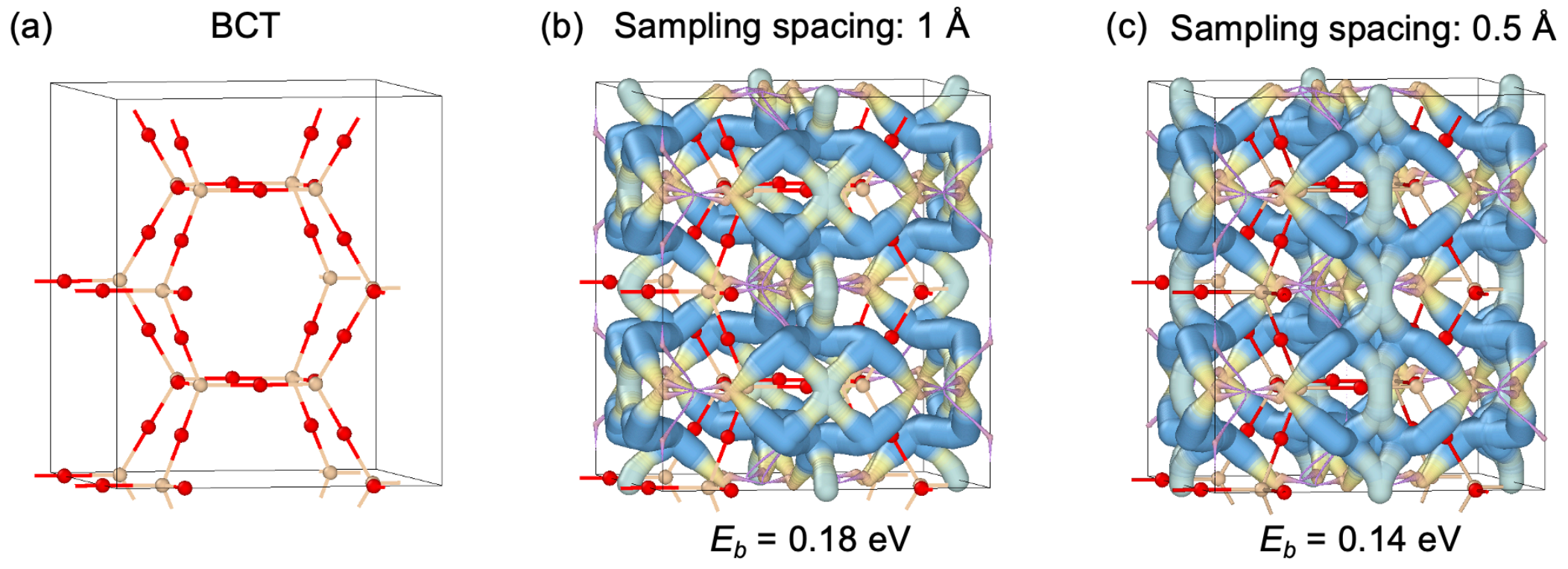

**Figure S1.** (a) Structure of the BCT zeolite. Diffusion networks and overall diffusion energy barriers $E_b$ of $Li^+$ based on a sampling spacing of (b) 1 Å and (c) 0.5 Å.

### (2) Symmetry tolerance for optimized adsorption sites in Step 4

After DFT optimization of the sampling points, we observe nearly overlapping adsorption sites. As shown in Fig. S2, the distances between some optimized adsorption sites in the ERI zeolites range from 0.22 to 0.46 Å, with energy differences of ~0.05 eV. To retain only one of the nearly overlapping sites, we set a symmetry tolerance of 0.5 Å in our calculations.

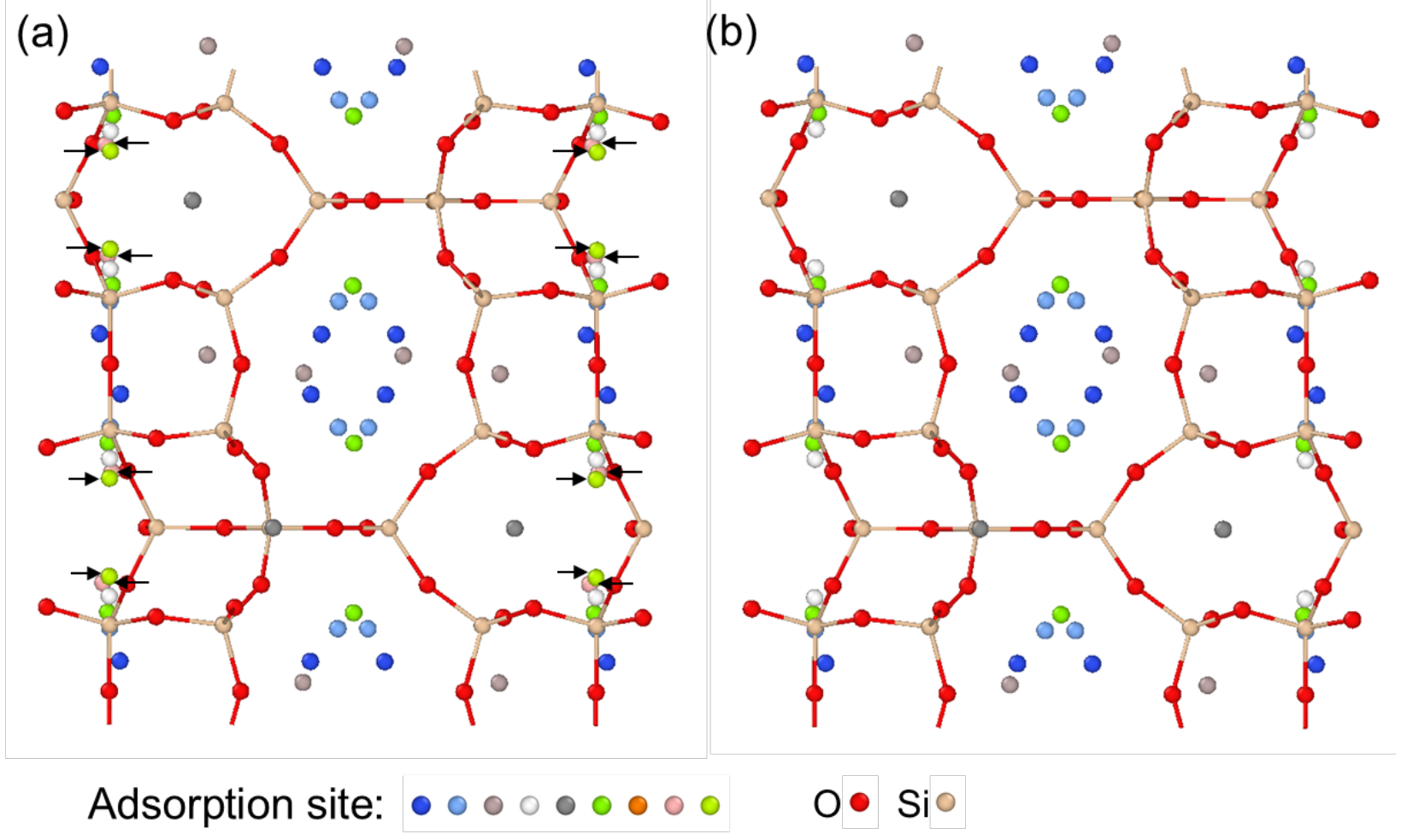

**Figure S2.** Adsorption sites of $Li^+$ in the ERI structures after optimization, (a) without using symmetry tolerance and (b) with a symmetry tolerance of 0.5 Å. Different color of spheres represent inequivalent stable adsorption sites. Black arrows in panel (a) indicate adsorption sites that are less than 0.5 Å from neighboring adsorption sites.

### (3) Reasonable hopping distance range in Step 5

To systematically evaluate potential hopping pathways, we consider all possible combinations of stable adsorption sites and then exclude those that are either too close or too distant. Given the symmetry tolerance of 0.5 Å used in Step 4, the minimum distance between stable adsorption sites is constrained to 0.5 Å. To determine the maximum hopping distance, we focus on the AHT zeolite, which features the largest twelve-membered Si-O rings among the studied zeolites and thus suggests a potential for long hopping distance. It turns out that the maximum distance between neighboring stable adsorption sites is 5.8 Å in AHT zeolite and there is no direct hopping path between them (Fig. S3). Consequently, we set the maximum hopping distance to 5 Å. With the above reasons, we consider hopping paths between adsorptions sites with distance ranging from 0.5 to 5 Å.

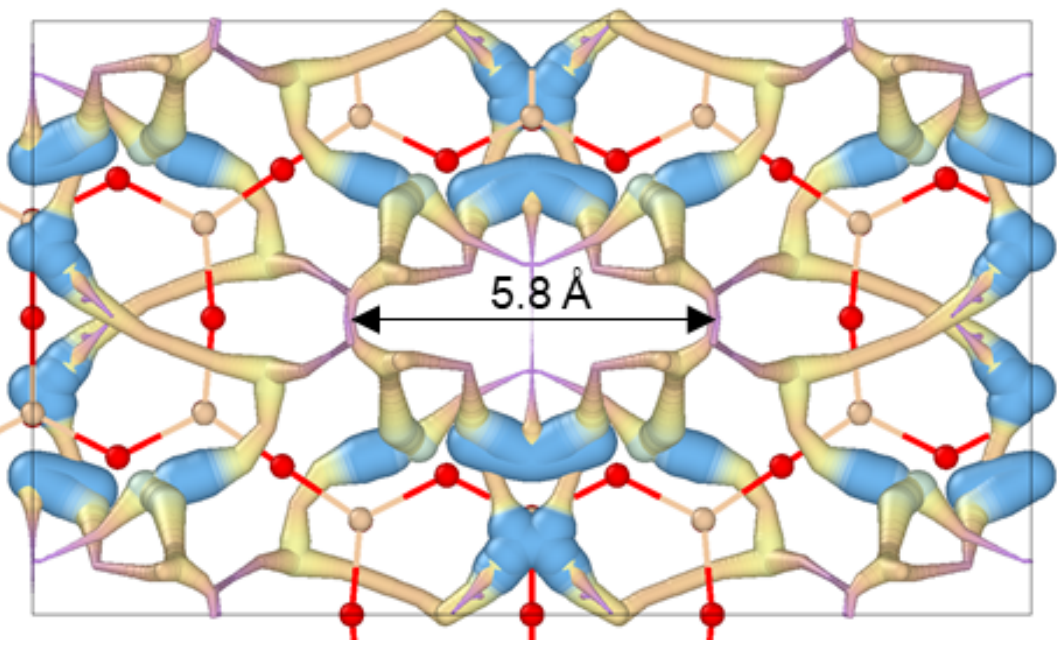

**Figure S3.** Diffusion network of the AHT zeolite and the maximum distances between neighboring stable adsorption sites.

### (4) Substituting hopping paths in Step 5

As illustrated in Fig. S4a, the $Li^+$ hopping within a six-membered Si-O ring of the ERI zeolite from state IS to state FS can occur through two inequivalent pathways. Path A follows a nearly straight line with a distance of 3.8 Å (IS → FS), while Path B is a curved route through three shorter ones: IS → 1 → 2 → FS, with respective distances of 1.5, 1.6, and 1.5 Å (Fig. S4b). Transition state calculations reveal an energy barrier of 1.0 eV for Path A and 0.7 eV for Path B (Fig. S4c). This finding is consistent with the fact that Path B effectively utilizes intermediate stable adsorption sites. For this reason, we substitute hopping paths if they can be replaced by shorter alternatives.

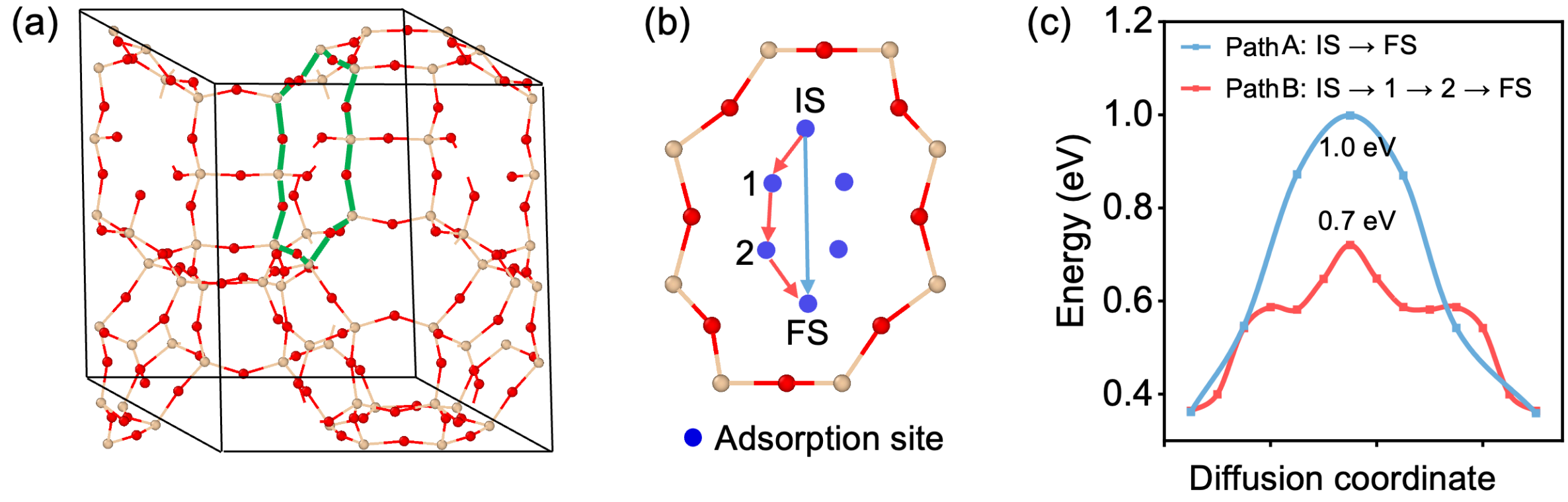

**Figure S4.** (a) ERI zeolite with a highlighted six-membered Si-O ring in green. (b) Two hopping pathways of $Li^+$ from state IS to state FS within the six-membered Si-O ring, and (c) the corresponding energy curves.

## III. Diffusion Energy Barriers and Networks of $Li^+$, $Na^+$, and $K^+$ in Examined Zeolites

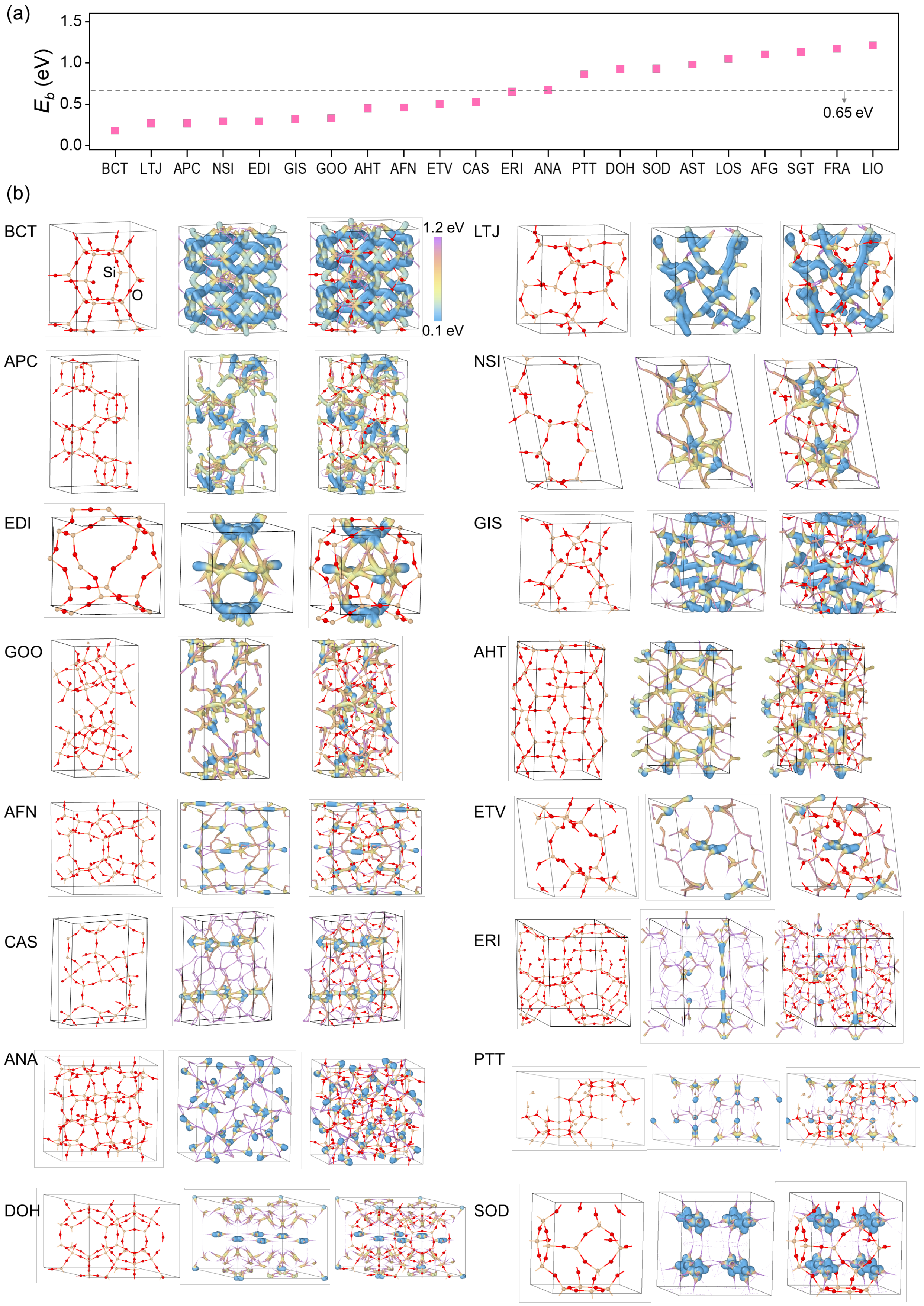

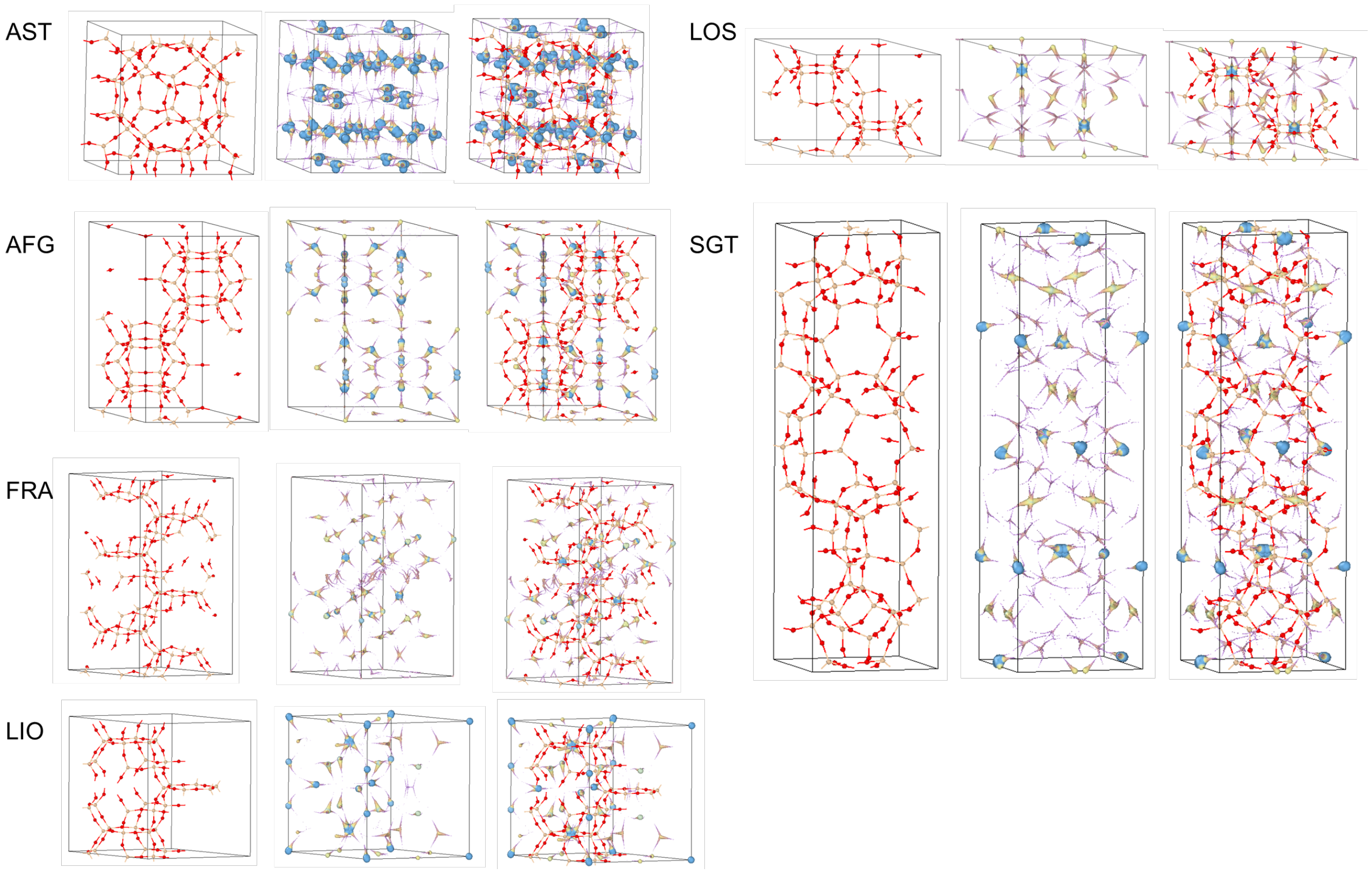

**Figure S5.** (a) Diffusion energy barrier and (b) diffusion networks of $Li^+$ in zeolites. The first to third columns depicts the zeolite structure, diffusion network, and their combination, respectively

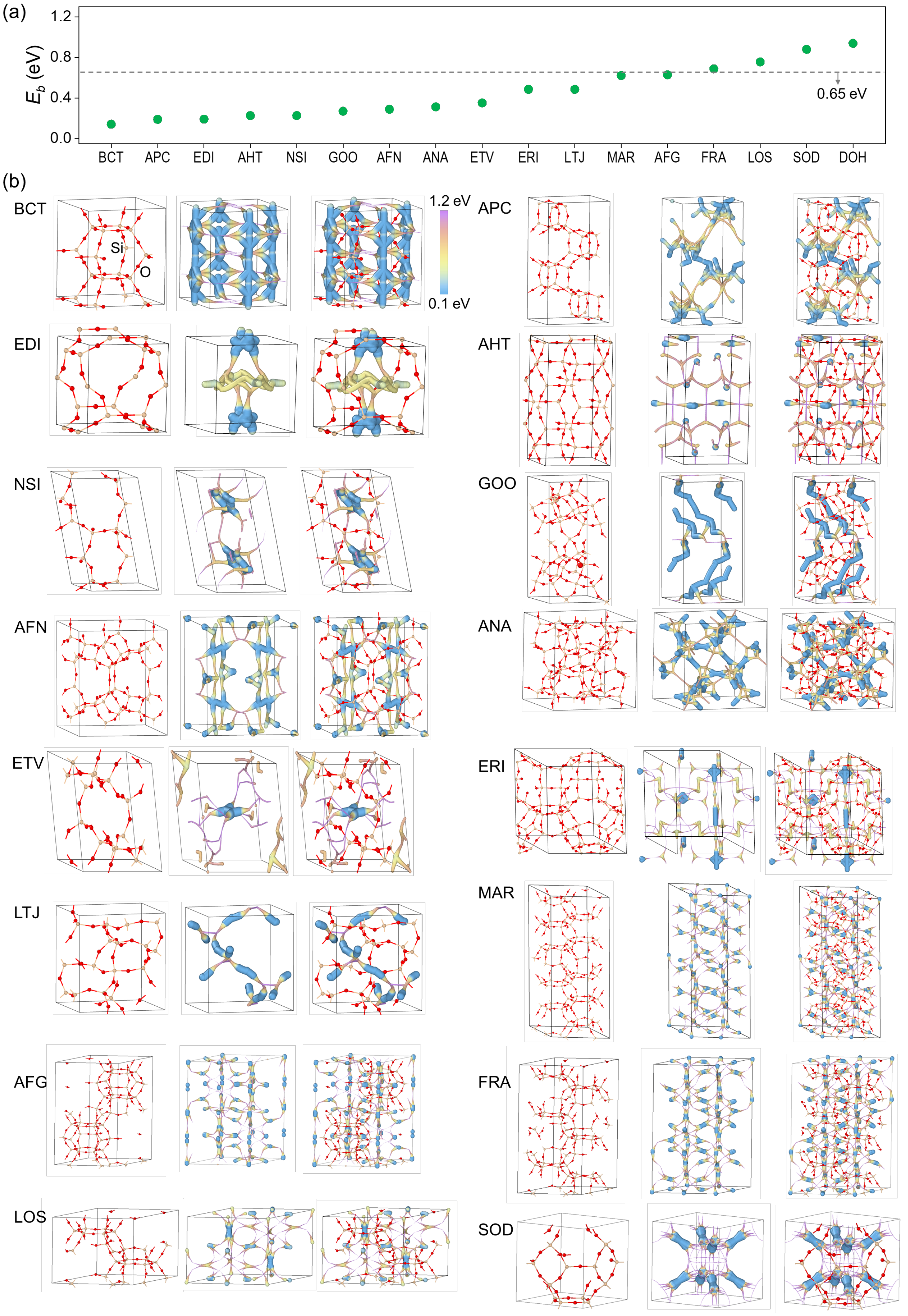

**Figure S6.** (a) Diffusion energy barrier and (b) diffusion networks of $Na^+$ in zeolites. The first to third columns depicts the zeolite structure, diffusion network, and their combination, respectively.

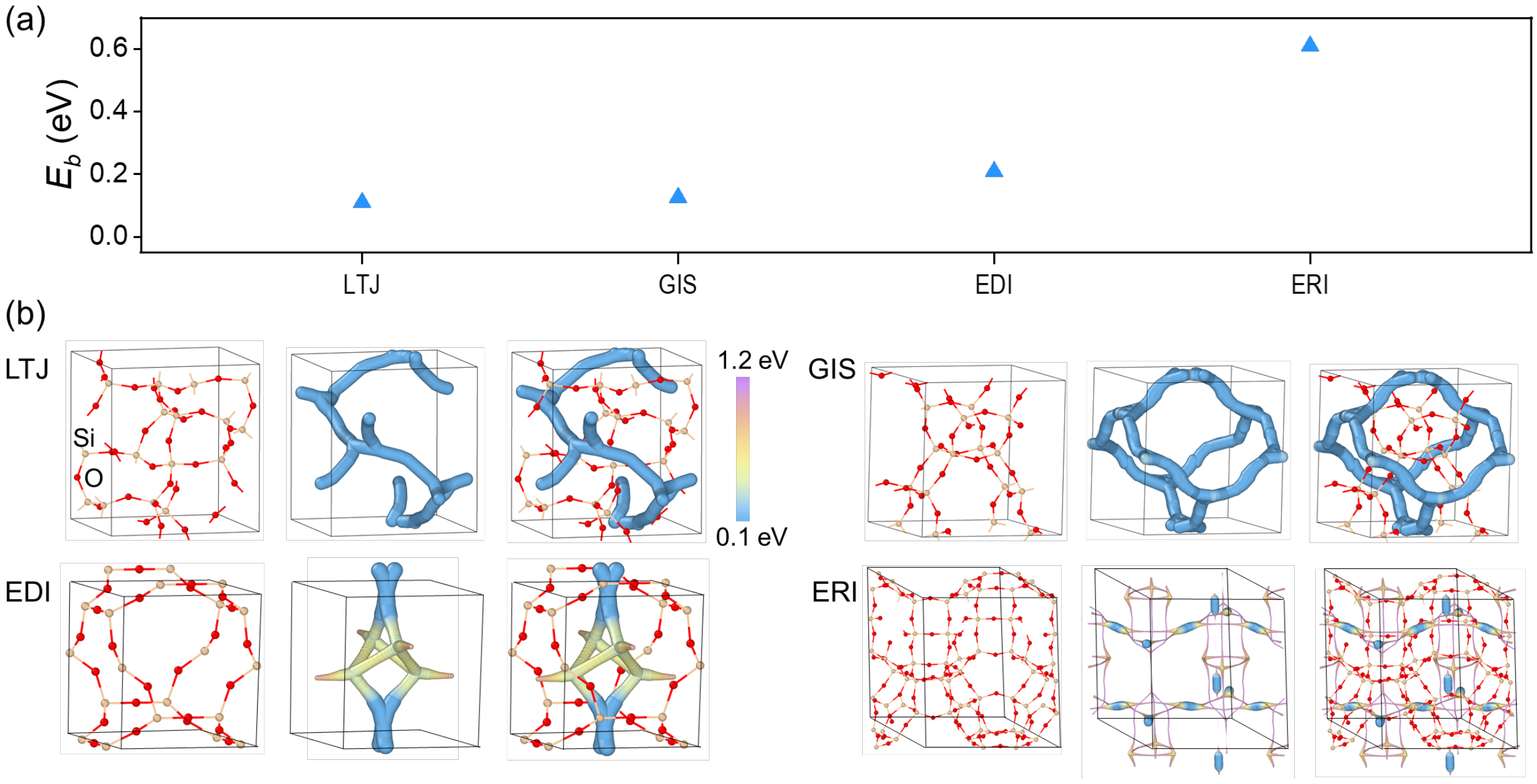

**Figure S7.** (a) Diffusion energy barrier and (b) diffusion networks of $K^+$ in zeolites. The first to third columns depicts the zeolite structure, diffusion network, and their combination, respectively.

**IV. Comparison of Adsorption Sites and Diffusion Pathways of $Li^+$, $Na^+$, and $K^+$ in Zeolite EDI**

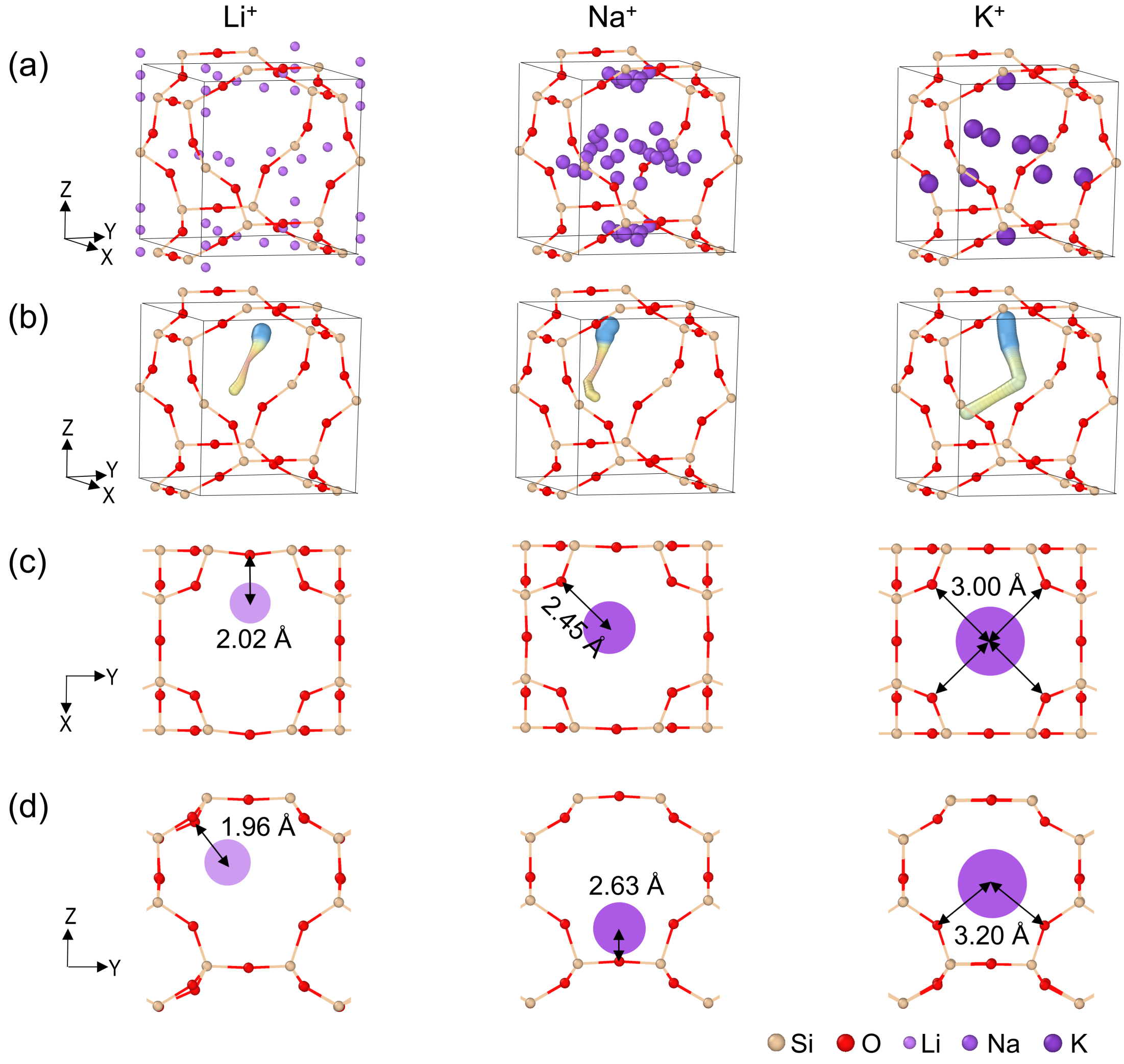

**Figure S8.** (a) Stable adsorption sites of $Li^+$, $Na^+$, and $K^+$ in EDI zeolite. (b) One diffusion path of $Li^+$, $Na^+$, and $K^+$ in EDI zeolite, and the corresponding (c) initial state and (d) final state. The sphere radius in panels (c) and (d) correspond to the ionic radius.

## V. Adsorption Energy of the Lowest Adsorption Site and That of the Transition State in Diffusion-Limited Step

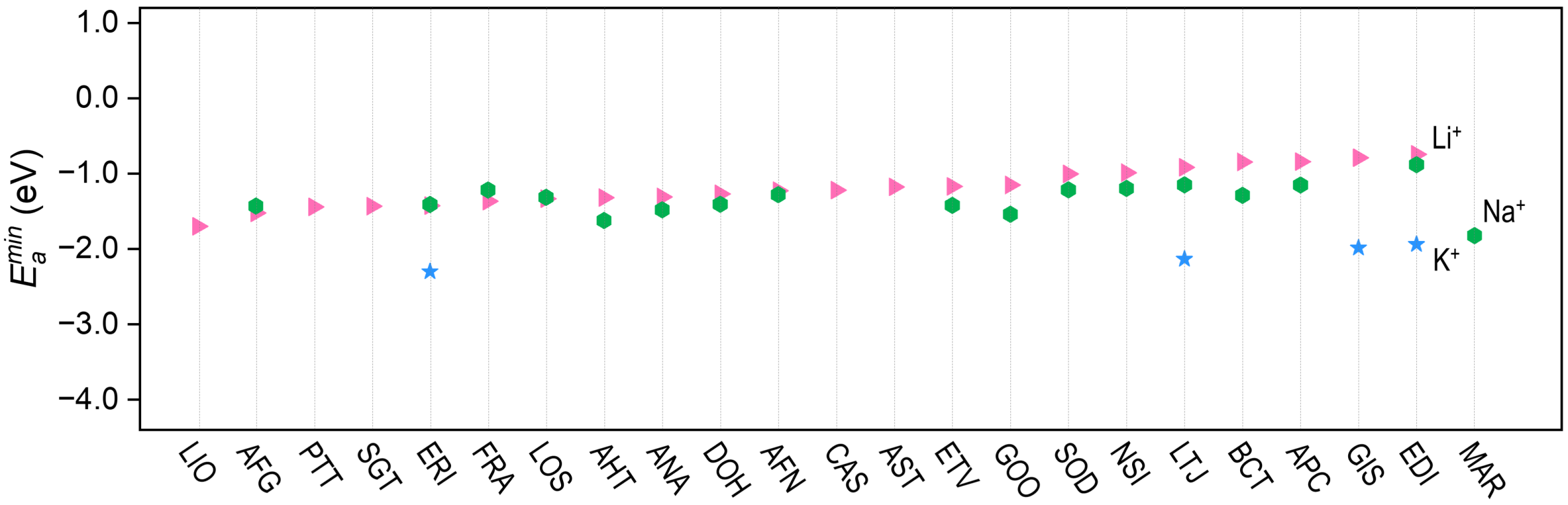

**Figure S9.** Lowest adsorption energies of $Li^+$, $Na^+$, and $K^+$ within all the examined zeolites.

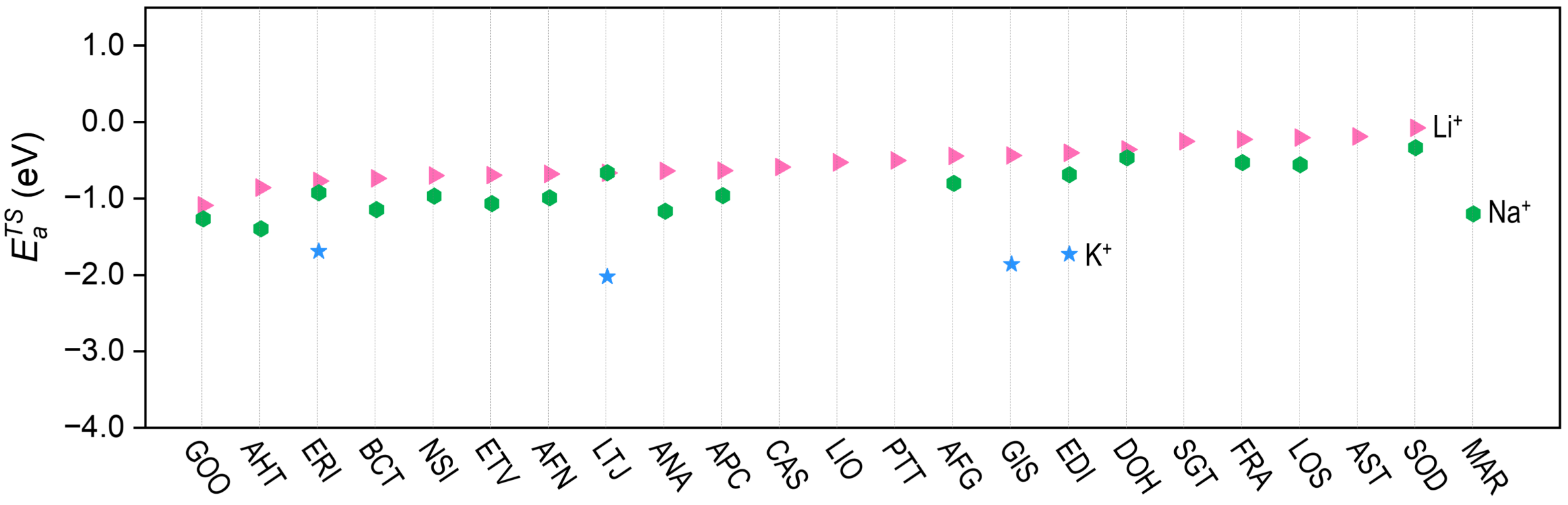

**Figure S10.** Adsorption energies of the transition state for $Li^+$, $Na^+$, and $K^+$ in the diffusion-limited step within all the examined zeolites.

## VI. Fitting between Adsorption Energies and Local Structural Features

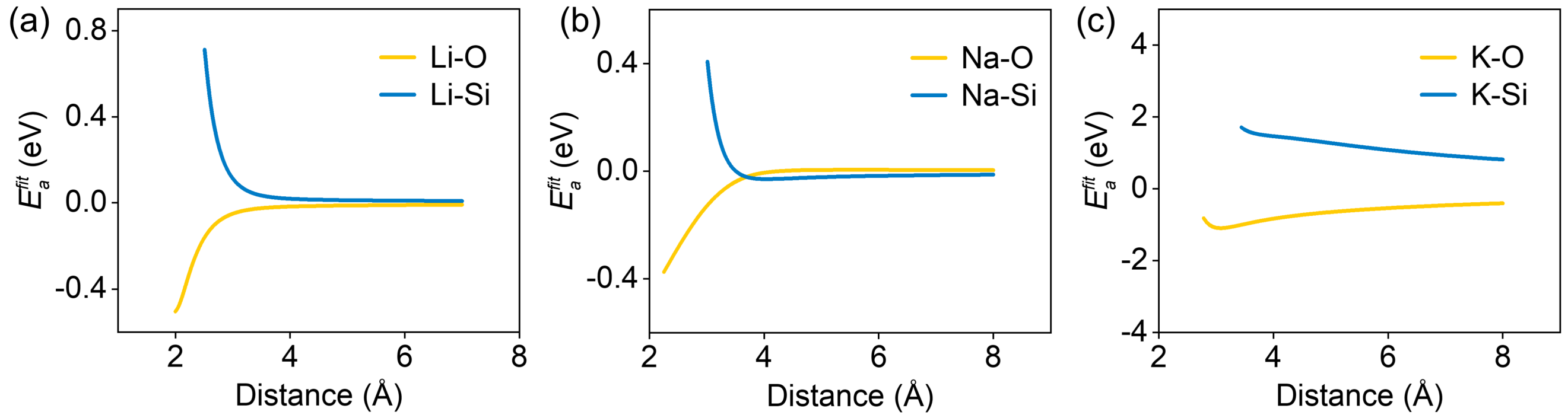

**Figure S11.** Fitting curves of adsorption energy with respect to distances of (a) Li-O and Li-Si, (b) Na-O and Na-Si, and (c) K-O and K-Si in all the examined zeolites.